\documentclass[aps,prl,reprint,groupedaddress,10pt,twocolumn]{revtex4}
\usepackage{graphicx}

\def\8{\infty}
\def\oh{\frac{1}{2}}

\def\undertext#1{\vtop{\hbox{#1}\kern 1pt \hrule}}

\def\be{\begin{equation}}
\def\ee{\end{equation}}
\def\bea{\begin{eqnarray} & &}
\def\eea{\end{eqnarray}}

\def\rfs#1{Eq.~(\ref{#1})}

\begin{document}
\title{Reentrant BCS-BEC crossover and a superfluid-insulator transition
  in optical lattices} 
\author{Zhaochuan Shen, L. Radzihovsky, V. Gurarie}
\affiliation{Department of Physics, University of Colorado, Boulder,
  CO 80309, USA}


\date{\today}

\begin{abstract}
  We study thermodynamics of a two-species Feshbach-resonant atomic
  Fermi gas in a periodic potential, focusing in a deep optical
  potential where a tight binding model is applicable.  We show that
  for more than half-filled band the gas exhibits a reentrant
  crossover with decreased detuning (increased attractive
  interaction), from a paired BCS superfluid to a Bose-Einstein
  condensate (BEC) of molecules of holes, back to the BCS superfluid,
  and finally to a conventional BEC of diatomic molecules. This
  behavior is associated with the non-monotonic dependence of the
  chemical potential on detuning and the concomitant
  Cooper-pair/molecular size, larger in the BCS and smaller in the BEC
  regimes.  For a single filled band we find a quantum phase
  transition from a band insulator to a BCS-BEC superfluid, and map
  out the corresponding phase diagram.
\end{abstract}


\maketitle
Feshbach resonances have become an essential experimental tool in the
exploration of interacting degenerate atomic gases, allowing, for a
realization of Fermi superfluidity tunable from a weakly-paired BCS to
a strongly-paired molecular
regime\cite{Jin,Ketterle,Thomas,GRaop,otherReviews,Chin:2010kl}.

Confinement of atomic gases in optical lattices is another powerful
technique for realizing and tuning strong
correlations\cite{Bloch:2008gl}, allowing experimental investigation
of a variety of lattice quantum many-body phenomena, such as the
superfluid to Mott insulator transition\cite{Greiner2002}.

Naturally, recent attention has focused on the rich combination of
Feshbach-resonant gases in optical lattices\cite{Kohl05,
  otherExp,Hulet1dFFLO}. Although considerable progress has recently
been made\cite{Zoller,Duan,HoHui,Sachdev,Buchler,FeiZhou,Stecher2011},
a general theoretical description of such system is challenging even
at the two-body level as it involves a projection of the Feshbach
resonance physics onto the eigenstates of the periodic
potential. Optical lattice shifts bulk Feshbach resonances and induces
new ones.

Yet it is possible to argue\cite{Hassan,SRGunpublished} that in a
sufficiently deep optical lattice (bandwidth much smaller than a
bandgap), even in the vicinity of a broad Feshbach resonance, the
single band tight binding model with on-site attraction (attractive
Hubbard model) $U(a)$, a function of a vacuum scattering length $a$ is
sufficient to describe the physics. Likewise, for a narrow Feshbach
resonance in a deep lattice a tight binding model of open-channel
fermions and closed-channel bosonic molecules coupled by an on-site
interconversion term, Eq.~(\ref{Hamiltonian1}) is appropriate.

\begin{figure}[htb]
\centering
\includegraphics[width=0.5\textwidth]{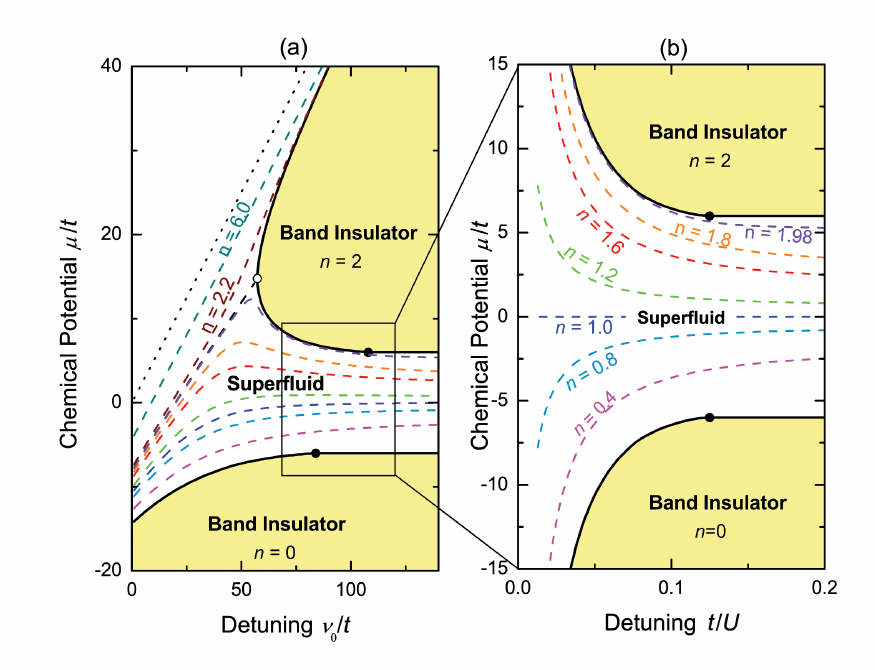}
\caption{The normalized chemical potential in, two-channel model (a)
 (for $g/t=20\sqrt{2}$) and one-channel model (b).  Colored dashed lines
  correspond to different density $n$ of the gas. The leftmost dashed
  line in (a) has $n>2$. The dots on black lines correspond to the
  threshold of the molecule formation in the vacuum.}
\label{picture1}
\end{figure}

\begin{figure}[htb]
\centering
\includegraphics[width=0.5\textwidth]{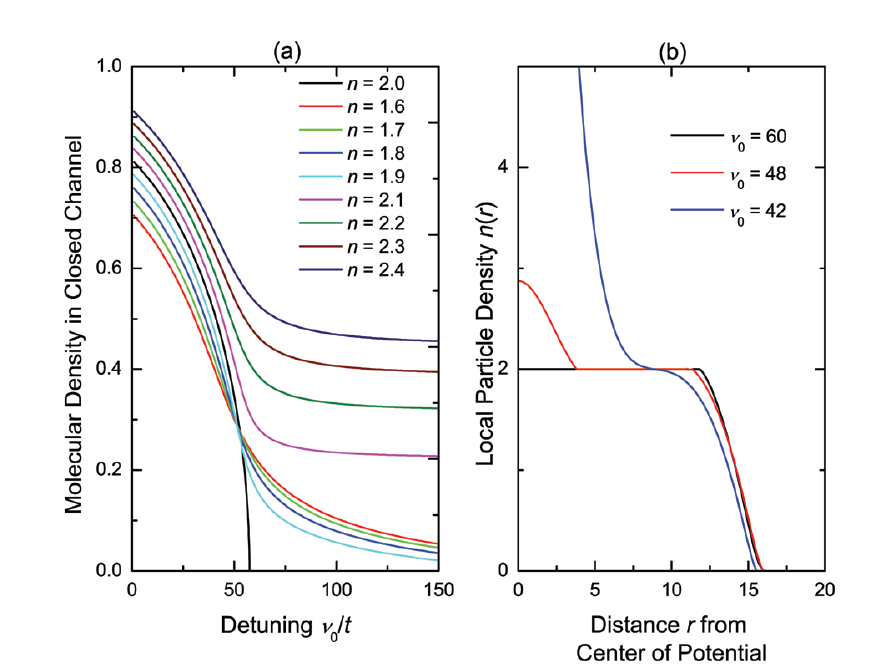}
\caption{ (a) Paired condensate (closed-channel molecular) density
  $|B|^2$ as a function of detuning at different total atom densities
  $n$. At total density $n=2$, the molecular density vanishes above
  certain critical detuning, signaling a quantum phase transition into
  a band insulator. (b) Radial atomic density $\rho(r)$ of a trapped
  gas in the local density approximation. The plateau corresponds to an
  $n=2$ band insulator shell, sandwiched by an inner ($n>2$) and
  outer ($n < 2$) superfluids.}
\label{phasetransition}
\end{figure}

In this Letter, focusing on the deep lattice regime we establish that
for a broad resonance, a two-component Fermi gas above half filling
undergoes a BCS-BEC crossover with the molecular condensate in the BEC
regime composed of holes; below half filling the crossover is to a
conventional condensate of diatomic molecules.  The BCS and BEC
regimes are separated by an analog of a unitary point where the
thermodynamic properties of the gas are given by a universal function
of the band filling fraction. For a narrow resonance
\cite{GRaop,Ohara2012}, while below half filling the behavior is
qualitatively equivalent to its broad-resonance counterpart, above
half filling the crossover is non-monotonic and reentrant. Upon
decreasing the detuning the phenomenology crosses over from a BCS
regime to a BEC regime of molecules composed of holes, back to a BCS
regime and finally to a BEC regime of diatomic molecules. This rich
behavior, illustrated in Fig.~(\ref{picture1}a) is associated with a
corresponding non-monotonic dependence of the chemical potential on
the Feshbach resonance detuning. Concomitantly, the size of a Cooper
pair/molecule changes non-monotonically, with the BCS and BEC regimes
respectively characterized by large and small pair sizes in units of
the interatomic spacing.  This resonant lattice phenomenology can in
principle be probed by measuring correlations in the gas after its
free expansion\cite{Jin2005,Altman}, through RF
spectroscopy\cite{GaeblerJin}, compressibility, and via details of the
atomic cloud's density profile\cite{Bloch:2008gl}, as illustrated in
Fig.~\ref{phasetransition}.

Finally, we find that for a narrow resonance, a fully filled lattice
(band filling of $2$) exhibits a quantum phase transition between a
band insulator and a paired superfluid, absent for a broad resonance
or in the absence of an optical lattice. This latter transition can be
best detected in the ``wedding cake'' density profile, a layered
structure of the gas when placed in an overall confining harmonic
potential, with a shell of a band insulator sandwiched by an inner
superfluid core and an outer superfluid shell.

We now outline the derivation of these predictions. We study fermionic
atoms in an optical lattice which contains $N$ sites and $M$ atoms
(atom filling fraction $n=M/N$) for both the one- and two-channel
models. The starting point is to consider the tight-binding
two-channel Hamiltonian
\begin{eqnarray} \label{Hamiltonian1}
 H &=& - t \sum_{\left<\mathbf{r},\mathbf{r}^\prime\right>,\sigma}c^\dagger_{\mathbf{r},\sigma}c_{\mathbf{r}^\prime,\sigma} - \mu \sum_{\mathbf{r},\sigma} c^\dagger_{\mathbf{r},\sigma} c_{\mathbf{r},\sigma} - t_b \sum_{\left<\mathbf{r},\mathbf{r}^\prime\right>} b^\dagger_\mathbf{r} b_{\mathbf{r}^\prime}\nonumber\\
 &+&(\nu_0 - 2 \mu) \sum_\mathbf{r} b^\dagger_\mathbf{r}b_\mathbf{r}+g \sum _\mathbf{r} \left(b^\dagger_\mathbf{r} c_{\mathbf{r},\uparrow} c_{\mathbf{r},\downarrow} + h. c.\right).
\end{eqnarray}
Here, $c^\dagger_{\mathbf{r},\sigma}$, $c_{\mathbf{r},\sigma}$ are the
creation and annihilation operators of the fermionic atoms at lattice
site $\mathbf{r}$ with spin $\sigma$, $b^\dagger_{\mathbf{r}}$,
$b_{\mathbf{r}}$ are creation and annihilation operators of bosonic
closed-channel molecules, $\mu$ is the chemical potential, $g$ is the
coupling, $\nu_0$ is the ``bare'' detuning, $t$ ($t_b$) is the hopping
matrix element for the atoms (molecules) and $\left<
  {\mathbf{r},\mathbf{r}'}\right>$ denotes pairs of nearest neighbor
sites.  In the absence of interactions, the bosons and fermions are
free particles, with the tight-binding dispersion of atoms given by
  \begin{equation} \epsilon_{\bf k} = -2 t (\cos k_x + \cos k_y + \cos
    k_z), \ \xi_{\bf k} = \epsilon_{\bf k} - \mu,
  \end{equation} 
  where the lattice constant is taken to be $1$.  Considerable
  progress in understanding the thermodynamics can be obtained through
  a mean-field approximation where closed-channel molecular field is
  replaced by a classical field:
\begin{equation}
b_\mathbf{q}\approx B \, \delta_{\mathbf{q},0},
\end{equation}
with $b_{\mathbf{q}}$ a Fourier transform of $b_{\bf r}$. 
Diagonalizing the resulting quadratic Hamiltonian and varying
corresponding ground state energy with respect to $\mu$ and $B^*$
gives the number and gap equations:
\begin{eqnarray}
 n &=& \int_{BZ}
 \frac{d^3k}{(2\pi)^3}\left(1-\frac{\xi_k}{\sqrt{\xi_k^2 + g^2 \left|
         B \right| ^2}}\right) + 2\left| B \right|^2
\label{particlenumber},\ \ \ \\
\nu_0 - 2 \mu
&=& \frac{g^2}{2}\int_{BZ} \frac{d^3 k}{(2\pi)^3} \frac{1}{\sqrt{\xi_k^2 +
    g^2 \left| B \right| ^2}},
\label{gap} 
\end{eqnarray}
with the integrals over the entire Brillouin zone (BZ). The solutions
give $\mu$ and $\left| B \right|$ as a function of detuning $\nu_0$.

We first analyze these equations in a limit of a broad resonance,
corresponding to taking $\nu_0$ and $g$ to infinity, while keeping
their ratio $U = g^2/\nu_0$ finite. In this limit, the closed-channel
molecules $b_{\mathbf{q}}$ can be adiabatically eliminated (integrated
out) reducing the Hamiltonian Eq.~(\ref{Hamiltonian1}) to that of an
attractive Hubbard model with the interaction strength $U$.

In this limit the Eqs.~(\ref{particlenumber}) and (\ref{gap}) become
\begin{eqnarray}
 n &  = & \int_{BZ}\frac{d^3 k}{(2\pi)^3}
\left( 1 - \frac{\xi_k }{\sqrt{\xi_k^2 + \Delta^2}}\right), 
\label{onechannelparticlenumber}\\ 
\frac{1}{U} & = &\frac 1 2 
\int_{BZ}\frac{d^3 k}{(2\pi)^3}\frac{1}{\sqrt{\xi_k ^2 + \Delta^2}}\label{onechannelgap} ,
\end{eqnarray} 
where $\Delta=g \left| B \right|$ is finite, while $\left|B \right|$
goes to zero in the broad resonance limit.

An important feature of these equations (by particle-hole symmetry
that holds even beyond the mean-field approximation \cite{Prokofiev})
is that the sign of the chemical potential $\mu$ depends on whether
the band filling fraction $n$ is above or below $1$ (corresponding to
above or below half filling), with $\mu=0$ for $n=1$ independent of
$\Delta$ and $U$.  Mathematically, the latter is captured by $\int
d^3k \, \xi_k F(\left| \xi_k \right|) = 0$ for any function $F(\left|
  \xi_k \right|)$ for $\mu=0$. To see that the sign of $\mu$ is
independent of $\Delta$ and therefore $U$, suppose $n>1$. Then for
$\Delta=0$ (corresponding to $U\rightarrow 0$), the right hand side of
Eq.~(\ref{onechannelparticlenumber}) reduces to the Fermi-Dirac
step-function, obviously giving $\mu>0$. Now, for a nonzero $\Delta$,
$\mu$ must remain positive since $\mu$ crossing zero at any $\Delta$
would imply $n=1$, contradicting the $n>1$ assumption. Similarly, if
$n<1$, then $\mu<0$, independent of $U$.

Solving Eqs.~(\ref{onechannelparticlenumber},\ref{onechannelgap})
numerically gives the relation between $\mu$ and $U$ shown in
Fig.~(\ref{picture1}b).  We thus observe that in this broad resonance
regime, for $n<1$ the system undergoes a conventional BCS-BEC
crossover, reaching the BEC regime for strong attractive interaction
$U$ (or equivalently for reduced detuning $\nu_0$), where large
negative chemical potential gives (negative) half the binding energy
of molecules formed by pairs of fermionic atoms.  In contrast, for
$n>1$, as a reflection of particle-hole mapping between $n<1$ and
$n>1$ fillings, the $\mu > 0$ chemical potential actually grows with
increased attractive interaction, crossing above the top of the band
for large attractive $U$. This corresponds to a BCS-BEC crossover to a
BEC of molecules of two holes in the Fermi sea with the chemical
potential tracking half of their binding energy.  At exactly half
filling ($n=1$) there is no BCS-BEC crossover, with chemical potential
remaining pinned exactly at $\mu=0$ for all $U$, consistent with
particle-hole symmetry of the Hubbard model at half-filling.

At $n=0$ and $n=2$ the
Eqs.~(\ref{onechannelparticlenumber},\ref{onechannelgap}) can be
straightforwardly solved analytically. For example, at $n=0$ we expect
that $\Delta=0$ and $\mu\le -6 t$, dropping below the bottom of the
band. Then Eq.~(\ref{onechannelparticlenumber}) is automatically
satisfied ($\xi_k > 0$ for all $\bf k$), and Eq.~(\ref{onechannelgap})
reduces to 
\be \label{eq:gapatlowdens} 
\frac{1}{U} - \frac{C}{4 t} =
\frac{1}{2} \int_{BZ} \frac{d^3k}{(2\pi)^3} \left[ \frac{1}{\xi_{\bf
      k}} - \frac{1}{\epsilon_{\bf k} + 6 t} \right], 
\ee 
where $C = \int_{BZ} \frac{d^3 k }{(2\pi)^3} \frac{1}{3 - \cos(k_x) -
  \cos(k_y)-\cos(k_z)} \approx 0.505$.
Writing the right hand side of the gap equation in this form allows
us to expand the cosines around ${\bf k}=0$ and extend the integral
to infinity (valid as long as $-\mu/(6 t) - 1 \ll 1$) to find
\begin{equation}\label{eq:umu}
\frac  1 U - \frac{C}{4 t}=   -  \frac{\sqrt{ |\mu| -6 t}}{8 \pi t^{3/2}}\label{muU}.
\end{equation}
This is the $\mu(U)$ relation at the lowest density, $n=0$,
corresponding to half of the binding energy of the zero momentum
molecule formed by two atoms. In order for the molecule to form in 3D,
$U$ must exceed a threshold attraction, $U>4t/C$, as is known in the
absence of lattice potential\cite{LLqm,GRaop}. We note, however, that
at a finite center of mass momentum at the Brillouin Zone boundary, a
molecule can form at an arbitrarily weak attraction even in
3D\cite{Stecher2011}.

Similarly by particle-hole symmetry, at $n=2$ close to the top of the
band $\mu(U)$ is also given by Eq.~(\ref{muU}).  The full $\mu(U,n)$
dependence for $0 < n < 2$ can be obtain through a straightforward
numerical or an approximate analytical analysis of
Eqs.~(\ref{onechannelparticlenumber}, \ref{onechannelgap}). However,
it should be kept in mind that these are in themselves only valid
within a mean field approximation, except at very small or very large
$U$ (at $n\not =1$).

We observe that at a special value $U=4t/C$ the interaction strength
drops out of Eqs.~(\ref{onechannelparticlenumber},
\ref{onechannelgap}), corresponding to a divergent scattering length,
just like at the unitary point of the BCS-BEC crossover in the absence
of a lattice potential. Although these mean-field equations are
approximate, this general feature holds more generally, with the
chemical potential at a unitary point $U=U_*$ given by a universal
function of particle density \be \mu/t = f(n).  \ee with the property
$f(2-n)=-f(n)$ dictated by the particle-hole symmetry.  In the dilute
$n \rightarrow 0$ limit, we recover the lattice-free result 
\be 
f(n)
\approx -6+ \xi \left( 3 \pi^2 n\right)^{2/3}, 
\ee 
where $\xi$ is the Bertsch parameter\cite{Veillette}, $\xi\approx
0.4$\cite{Carlson,Stewart,Carlson2011}. The full function $f(n)$ is
not known but can be computed numerically.

In contrast to the above broad resonance limit (one-channel model),
where $n$ is restricted to $0\leq n \leq 2$, in the two-channel model,
\rfs{particlenumber} the filling fraction is no longer limited by two
particles per site, as closed-channel bosonic states can accommodate
an arbitrary number of fermionic atom pairs even if all states in the
fermionic band are occupied. We next study the ground state behavior
of the two-channel model encoded in
Eqs.~(\ref{particlenumber},\ref{gap}), that in the narrow resonance
limit, $g\sqrt{n}\ll E_F$ are quantitatively trustworthy across the
entire phase diagram and in the opposite broad resonance limit reduce
to those of the one-channel model discussed above.

Solving these equations numerically leads to the $\mu-\nu_0$ phase
diagram illustrated in Fig.~(\ref{picture1}a). Its many features can
be understood analytically, particularly for a narrow resonance. It
displays a band insulator (BI) and a paired superfluid (SF) phases,
depending on the range of the chemical potential and detuning. For
large positive detuning $\nu_0$, closed-channel molecules are
separated by a large gap above the fermionic band, leading to a weak
attractive interaction for the atoms.  Thus, for a partially filled
fermionic band, $0 < n < 2$, the chemical potential $\mu(n)$ sits
within the band $-6t < \mu < 6t$, and the ground state is a
weakly-paired BCS superfluid. Increasing the filling to $n=2^-$ pushes
the chemical potential to the top of the band, $\mu(2^-)=6t$. Since
the fermionic band is then full at $n=2$, a further increase in $n$
can only be accommodated by populating the closed-channel molecular
state. For large $\nu_0$, the chemical potential therefore jumps from
$\mu(2^-)=6t$ to $\mu(2^+)\approx\nu_0/2$, which thus determines the
lower and upper phase boundaries of the $n=2$ BI.

Reducing the detuning $\nu_0$ brings down the molecular state and
leads to its hybridization with the pairs of the fermionic band states
and a concomitant increases in the attractive interactions. Below
half-filling, $n < 1$ this leads to a monotonic emptying of the
fermionic band as the BCS superfluid crosses over to the molecular
BEC, familiar from a narrow resonance BCS-BEC crossover in the absence
of the periodic potential\cite{GRaop}.

The phase boundary between the paired superfluid and $n=0$ BI (vacuum)
can then be found exactly, as it corresponds to a limit of a two-atom
ground state, with $\mu_c(\nu_0)\equiv\mu(\nu_0,n=0,B=0)$. While for
large positive $\nu_0$, the phase boundary $\mu_c(\nu_0) = -6t$
follows the bottom of the band, for $\nu_0 < \nu_0^*$, a true stable
molecular bound state (not just a resonance) peels off from the bottom
of the band following half of the molecular binding energy. To see
this emerge from Eqs.~(\ref{particlenumber},\ref{gap}), we set $n=0$,
$B=0$ and note that for $\mu\leq -6t$ (below the bottom of the band)
the number equation, \rfs{particlenumber} is automatically
satisfied. The gap equation then gives
\begin{equation} \nu_0 = \frac{g^2}{4 t} C + 2 \mu - g^2
  \frac{\sqrt{|\mu| -6t}}{8 \pi t^{3/2}}\label{boundary}.
\end{equation}
For detuning just below $\nu_0^*=g^2C/(4t)-12t$ the dependence is
quadratic $\mu_c(\nu_0)+6t\sim - (\nu_0^*-\nu_0)^2$ and crosses over
to linear behavior $\mu_c(\nu_0)\sim\oh(\nu_0-\nu_0^*)$, following the
closed-channel level, as expected from the lattice-free
analysis\cite{GRaop}.

Similar to broad resonance, Eq.~(\ref{boundary}) also depicts the
phase boundary of $n=2$. On this boundary the threshold value of
detuning below which the molecular bound state of two holes can first
appear is given by $\nu_0^*=g^2C/(4t)+12t$, with $\mu_c=6 t$,
indicated by a dot in Fig.~(\ref{picture1}a).  Near and below this
threshold point, $\sqrt{\mu-6 t }$ dominates and leads to a lower
branch that grows quadratically, corresponding to half the binding
energy of two holes. This reflects the particle-hole symmetry near
this point and is consistent with earlier analysis of the broad
resonance limit. The upper branch anticipated on general grounds, at
large positive detuning asymptotes to the linearly growing solution,
$\mu_c(\nu_0)\sim\nu_0/2$, corresponding to the nearly decoupled
closed-channel state that forms the upper phase boundary of the $n=2$
BI.

From further analysis of Eqs.~(\ref{particlenumber},\ref{gap}), for
$n\neq 2$ we find the ground state is a superfluid for all $\nu_0$,
with the chemical potential $\mu(n,\nu_0)$ a smooth function of the
filling $n$ and detuning $\nu_0$ (i.e., exhibits a finite superfluid
compressibility), as indicated by dashed curves in
Fig.~(\ref{picture1}a). For $1 < n < 2$, $\mu(n,\nu_0)$ displays a
non-monotonic dependence with $\nu_0$, that leads to a reentrant
BCS-BEC crossover of holes at intermediate detuning and a conventional
one of atoms at large negative detuning.

In contrast, at $n=2$ the system undergoes a quantum phase transition,
shown on Fig.~(\ref{phasetransition}a), from a band insulator to a
paired-hole superfluid as $\nu_0$ is lowered below a critical value
$\nu_c=\nu_0^*- \frac{g^4}{512 \pi^2 t^3}$ corresponding to the tip of
the band insulator lobe at $\mu_c\equiv\mu(2,\nu_c)=6t +
\frac{g^4}{1024\pi^2 t^3}$ in Fig.~(\ref{picture1}a).

The onset of superfluid order close $\nu_c$, where $B$ is small, can
be studied analytically by expanding the number and gap equations,
Eqs.~(\ref{particlenumber},\ref{gap}) in Taylor series in $\left|B
\right|^2$.  This gives that inside the SF phase below $\nu_c$ at
$n=2$, $\mu(2,\nu_0)/t$ is a line of slope $3/4$. Around the critical
point, the superfluid order parameter displays the standard mean-field
onset, $B\sim (\nu_c-\nu_0)^{1/2}$.  The paired condensate (encoded in
closed-channel molecular) density $\left| B \right|^2$ as a function
of detuning, for different atom densities $n$ is illustrated in
Fig.~(\ref{phasetransition}a), clearly revealing the BI-SF at $n=2$.

Because the fermionic $n$=2 BI is continuously connected to a bosonic
Mott insulator at filling of one boson per site, beyond mean-field
theory (valid for narrow resonance) we expect this particle-hole
symmetric transition to be in the $d+1$ dimensional $xy$ university
class. Among a variety of other probes, such as thermodynamics,
density profile, time of flight, and noise, the BI-SF transition can
be detected via compressibility, which enters the speed of sound for
the superfluid mode, that we show vanishes at the critical point and
grows as $\left(n-2 \right)^{2/3}$ away from it\cite{SRGunpublished}.

To make contact with cold-atoms experiments, the trap potential $V(r)$
must be incorporated. This can be straightforwardly done through the
local density approximation, $\mu\rightarrow\mu_{\rm local}(r)=\mu -
V(r)$, where the local chemical potential $\mu_{\rm local}(r,N,\nu_0)$
is maximum in the center of the cloud and drops off at its edges, and
the global $\mu(N,\nu_0)$ is set by the total number of atoms $N$ in
the gas. The resulting radial density profile $\rho(r)$ is simply
determined by a cut through the bulk phase diagram in
Fig.~(\ref{picture1}a) with $\rho(r)=n(\mu_{\rm local}(r,N,\nu_0))$.
For $\nu_0 > \nu_c$ and average filling above $2$ this predicts an
$n>2$ superfluid core, surrounded by a shell of an $n=2$ band
insulator (with requisite ``wedding cake'' plateau), that is further
surrounded by a superfluid at $n<2$.  This is shown in
Fig.~(\ref{phasetransition}b) for different detunings.

To summarize, we studied an s-wave Feshbach resonant Fermi gas in a
deep lattice potential faithfully modeled by a single band two-channel
model. We showed that for above half lattice filling it exhibits an
interesting reentrant BCS-BEC crossover phenomenology of paired holes
and atoms associated with the nonmonotonic dependence of the chemical
potential on detuning. For a single filled band we find a quantum
phase transition from an $n=2$ band insulator to a BCS-BEC superfluid,
and map out the corresponding phase diagram. We expect that these
predictions should be testable with current state of the art
experiments on Feshbach-resonant Fermi gases in optical lattices.

We are grateful to N. Andrei for the early suggestions and discussions concerning  the  one dimensional version of the problem considered here. 
We acknowledge support by the NSF, through PHY-0904017 (ZS,VG) and 
through DMR-1001240 (LR).

\bibliography{ref.bib}
\end{document}